\documentclass[twocolumn,aps,pra,showpacs]{revtex4}
\usepackage[T1]{fontenc}
\usepackage[latin9]{inputenc}
\usepackage{amstext}
\usepackage{graphicx}

\makeatletter

\newcommand{\lyxdot}{.}

\@ifundefined{textcolor}{}
{%
 \definecolor{BLACK}{gray}{0}
 \definecolor{WHITE}{gray}{1}
 \definecolor{RED}{rgb}{1,0,0}
 \definecolor{GREEN}{rgb}{0,1,0}
 \definecolor{BLUE}{rgb}{0,0,1}
 \definecolor{CYAN}{cmyk}{1,0,0,0}
 \definecolor{MAGENTA}{cmyk}{0,1,0,0}
 \definecolor{YELLOW}{cmyk}{0,0,1,0}
}

\makeatother

\begin{document}

\title{Bose-Bose Mixtures with Synthetic Spin-Orbit Coupling in Optical
Lattices}

\author{Liang He$^{1,3}$, Anchun Ji$^{2}$, and Walter Hofstetter$^{1}$}

\affiliation{$^{1}$Institut für Theoretische Physik, Goethe\textendash{}Universität,
60438 Frankfurt/Main, Germany}

\affiliation{$^{2}$Center for Theoretical Physics, Department of Physics, Capital
Normal University, Beijing 100048, China}

\affiliation{$^{3}$Institut für Theoretische Physik, Leopold-Franzens Universität
Innsbruck, A-6020 Innsbruck, Austria}
\begin{abstract}
We investigate the ground state properties of Bose-Bose mixtures with
Rashba-type spin-orbit (SO) coupling in a square lattice. The system
displays rich physics from the deep Mott-insulator (MI) all the way
to the superfluid (SF) regime. In the deep MI regime, novel spin-ordered
phases arise due to the effective Dzyaloshinskii-Moriya type super-exchange
interactions. By employing the non-perturbative Bosonic Dynamical
Mean-Field-Theory (BDMFT), we numerically study and establish the
stability of these magnetic phases against increasing hopping amplitude.
We show that as hopping is increased across the MI to SF transition,
exotic superfluid phases with magnetic textures emerge. In particular,
we identify a new spin-spiral magnetic texture with spatial period
$3$ in the superfluid close to the MI-SF transition. 
\end{abstract}


\maketitle

\section{Introduction}

Spin-orbit (SO) effects have been known in solid-state physics for
a long time and have for example been observed in graphite more than
four decades ago \cite{SOC_in_graphite}. But their main effects were
believed to simply split the spin degenerate band structures \cite{Ashcroft_Mermin_book}.
It is only until recently that both theory and experiment have revealed
that strong spin-orbit coupling can give rise to dramatic qualitative
effects, leading to distinct and novel phases of matter such as topological
band insulators \cite{Hasan_Kane_RMP,Qi_Zhang_RMP}. This has motivated
experimental developments in the field of quantum simulation with
ultracold atoms, where tunable SO coupling and more generally synthetic
non-abelian gauge fields are created in both neutral bosonic atoms
\cite{SGF_bos_raman_1,SGF_bos_raman_2,SGF_bos_raman_3} and fermionic
atoms \cite{SGF_fer_raman_1,SGF_fer_raman_2} via Raman processes
or by driven optical lattices \cite{SGF_bos_driven_lattice}.

This progress has stimulated interesting studies on the physics of
SO coupled Bose-Bose mixtures subjected to an optical lattice \cite{SOC_BOS_lat_0,SOC_BOS_lat_1,SOC_BOS_lat_2,SOC_BOS_lat_3,SOC_BOS_lat_4,SOC_BOS_lat_5},
where the Mott insulator to superfluid phase transition \cite{SOC_BOS_lat_0,SOC_BOS_lat_1},
magnetic order in the deep Mott insulator regime \cite{SOC_BOS_lat_1,SOC_BOS_lat_2,SOC_BOS_lat_3,SOC_BOS_lat_4},
and superfluid phases \cite{SOC_BOS_lat_0,SOC_BOS_lat_1,SOC_BOS_lat_3,SOC_BOS_lat_5}
are investigated. In particular, in the strongly interacting regime,
a super-exchange spin model with Dzyaloshinskii-Moriya (DM) type interactions
\cite{Dzaloshinsky_paper,Moriya_paper} can be derived by second-order
perturbation theory. And various exotic spin-textures are predicted
by classical Monte-Carlo simulations or spin-wave analysis of the
effective spin model \cite{SOC_BOS_lat_1,SOC_BOS_lat_2,SOC_BOS_lat_3,SOC_BOS_lat_4}.
Similar spin textures induced by the DM-type interactions have also
been found in various solid state materials and attracted great interest
\cite{Pesin,Heinze,Banerjee}. Moreover, it is interesting to notice
that bosons in a certain class of frustrated lattices have similar
dispersion relations as particles with isotropic Rashba SO coupling
\cite{resembling_soc_coupling}, which could indicate that similar
exotic magnetic phases may also be relevant in these systems. 

However, there remain fundamental open questions. Generally, the second-order
super-exchange model only applies in the deep Mott regime, not to
mention certain cases, e.g. in presence of geometric frustration,
where it does \emph{not} even apply in the deep Mott regime \cite{bos_mix_tri_lat}.
One highly relevant question for experiments aiming at direct observation
of these exotic spin-textures at realistic temperature is whether
they are stable against increasing hopping amplitude, in particular
in the Mott regime close to the Mott insulator to superfluid (MI-SF)
phase transition, where the second-order super-exchange model no longer
holds true. Another important issue is that the two-particle interactions
play a fundamental role in determining the properties of the superfluid
phase in the vicinity of the MI-SF phase transition, where the spin-orbit
coupling can induce exotic superfluid phases.

In this work, we investigate these fundamental issues within the non-perturbative
theoretical framework of Bosonic Dynamical Mean Field Theory (BDMFT)
\cite{BDMFT_1,BDMFT_2,BDMFT_3,BDMFT_4}. We find that the exotic spin-textures
are robust throughout the whole Mott regime, see Fig.~\ref{Flo:phase_digram_alpha_t}.
This is of practical importance for experimental observation of these
phases at realistic temperatures. In the vicinity of the MI-SF transition,
we show that exotic superfluid phases arise as a result of the interplay
between onsite interaction and SO coupling, as shown in Fig.~\ref{Flo:Magnetic_SF_configuration}.
In particular, we identify a new spin-spiral magnetic texture with
spatial period $3$ on the SF side near the MI-SF transition, see
Fig.~\ref{Flo:Magnetic_SF_configuration}.b1.

\section{Model And Method}

We consider two species of bosons with SO coupling loaded in a square
optical lattice. For sufficiently low filling, we model the system
by a two-component Bose-Hubbard model in the lowest band approximation,
which reads 
\begin{equation}
H=-t\sum_{\langle i,j\rangle}(\psi_{i}^{\dagger}\mathcal{R}_{ij}\psi_{j}+h.c.)+\frac{1}{2}\sum_{i\sigma\sigma'}U_{\sigma\sigma'}\hat{a}_{i\sigma}^{\dagger}\hat{a}_{i\sigma'}^{\dagger}\hat{a}_{i\sigma'}\hat{a}_{i\sigma}.\label{eq:Hamiltonian}
\end{equation}
Here $\langle i,j\rangle$ denotes nearest-neighbor sites, $\psi_{i}^{\dagger}\equiv(\hat{a}_{i\uparrow}^{\dagger},\hat{a}_{i\downarrow}^{\dagger})$
with $\hat{a}_{i\sigma}^{\dagger}$($\hat{a}_{i\sigma}$) being bosonic
creation (annihilation) operators of the two species labeled by pseudo-spin
$\sigma=\uparrow,\downarrow$, on site $i$ in the Wannier representation.
$t$ is the overall hopping amplitude and the matrix $\mathcal{R}_{ij}=\exp[i\vec{A}\cdot(\mathbf{r}_{i}-\mathbf{r}_{j})]\,,$
where $\vec{A}=(\alpha\sigma_{y},\beta\sigma_{x},0)$ denotes a static,
non-abelian gauge field which can be generated by using a two-photon
Raman process \cite{Proposal_Rashba_SOC}. Here, $\sigma_{x},\sigma_{y}$
and $\sigma_{z}$ are the Pauli matrices. We shall consider the case
$\beta=-\alpha$, which implies that the SO coupling is of Rashba
type. In this case, the explicit form of $\mathcal{R}_{ij}$ reads:
$\mathcal{R}_{ij}=\cos\alpha\cdot\mathbf{I}\pm i\sin\alpha\sigma_{y},\textnormal{ for }\mathbf{r}_{i}-\mathbf{r}_{j}=\pm\hat{\mathbf{x}}$
and $\mathcal{R}_{ij}=\cos\alpha\cdot\mathbf{I}\mp i\sin\alpha\sigma_{x},\textnormal{ for }\mathbf{r}_{i}-\mathbf{r}_{j}=\pm\hat{\mathbf{y}}.$
We remark here that a similar type of SO coupling, which is a mixture
of Rashba and Dresselhaus coupling, has been recently realized experimentally
in trapped two-species bosonic quantum gases \cite{SGF_bos_raman_3}.
The second term in Eq.~(\ref{eq:Hamiltonian}) describes the on-site
interactions. We choose the intra-species repulsion $U_{\downarrow\downarrow}=U_{\uparrow\uparrow}\equiv V$,
and set the inter-species interaction $U_{\uparrow\downarrow}=U_{\downarrow\uparrow}\equiv U=\lambda V$
with a dimensionless parameter $\lambda$. In our investigation, we
shall focus on the ground state properties of the system at total
filling per site $\rho\equiv\sum_{i}\langle\hat{a}_{i\uparrow}^{\dagger}\hat{a}_{i\uparrow}+\hat{a}_{i\downarrow}^{\dagger}\hat{a}_{i\downarrow}\rangle/N_{\mathrm{lat}}=1$,
with $N_{\mathrm{lat}}$ being the number of lattice sites.

To investigate the properties of the system in the full range of interactions,
i.e. from the strong coupling deep MI regime all the way to the SF
at weak coupling, we apply BDMFT \cite{BDMFT_1,BDMFT_2,BDMFT_3,BDMFT_4},
which is non-perturbative and can capture the local quantum fluctuations
exactly. For exploring various possible exotic magnetic or superfluid
phases which break the translational symmetry of the lattice, here
we specifically employ real-space BDMFT (R-BDMFT) \cite{R-BDMFT},
which generalizes BDMFT to a position-dependent self-energy and captures
inhomogeneous quantum phases. Within this approach, the physics on
each lattice site is determined from a local effective action obtained
by integrating out all other degrees of freedom in the lattice model,
excluding the lattice site considered. The local effective action
is then represented by an Anderson impurity model \cite{BDMFT_1,BDMFT_2,BDMFT_3,BDMFT_4}.
We use exact diagonalization (ED) \cite{ED_impurity_solver_1,ED_impurity_solver_2}
of the effective Anderson Hamiltonian with a finite number of bath
orbitals to solve the local action. For the results presented in this
work, $n_{\mathrm{bath}}=4$ bath orbitals and a $12\times12$ lattice
are chosen. Typical physical results are checked with a larger number
of bath orbitals ($n_{\mathrm{bath}}=5$) and lattice size ($24\times24$)
and no qualitative difference are found. Details of the R-BDMFT method
can be found in previous work \cite{R-BDMFT}.

\section{Results}

\subsection{Mott-insulator regime}

In the deep MI regime for $U_{\sigma\sigma'}\gg t$, the physics of
the system can be captured by an effective spin model obtained by
second order perturbation theory \cite{SOC_BOS_lat_1,SOC_BOS_lat_2,SOC_BOS_lat_3,SOC_BOS_lat_4},
which is given by 
\begin{equation}
H_{\textnormal{eff}}=\sum_{i,\delta=\hat{\mathbf{x}},\hat{\mathbf{y}}}\left[\sum_{a=x,y,z}J_{\delta}^{a}S_{i}^{a}S_{i+\delta}^{a}+\mathbf{D_{\delta}}\cdot(\mathbf{S}_{i}\times\mathbf{S}_{i+\delta})\right],\label{eq:H_eff}
\end{equation}
where $S_{i}^{a}\equiv\psi_{i}^{\dagger}\sigma^{a}\psi_{i}/2$ with
$a=x,y,z$. The first and the second term denote the Heisenberg-type
(H-type) and Dzyaloshinskii-Moriya type (DM-type) super-exchange coupling
respectively. $J_{\delta}^{a}$ and $\mathbf{D_{\delta}}$ are the
corresponding coupling strengths. The explicit form of $J_{\delta}^{a}$
and $\mathbf{D_{\delta}}$ are given by $J_{\hat{x}}^{x}=J_{\hat{y}}^{y}=-4t^{2}\cos(2\alpha)/\lambda V$,
$J_{\hat{x}}^{y}=J_{\hat{y}}^{x}=-4t^{2}/\lambda V$, $ $$J_{\hat{x}}^{z}=J_{\hat{y}}^{z}=-4t^{2}(2\lambda-1)\cos(2\alpha)/\lambda V,$
$\mathbf{D}_{\hat{x}}=-4t^{2}\sin(2\alpha)\hat{y}/\lambda V$ and
$\mathbf{D}_{\hat{y}}=4t^{2}\sin(2\alpha)\hat{x}/\lambda V$. Since
these two types of couplings favor qualitatively different types of
magnetic orders, namely, H-type couplings favor either ferromagnetic
or anti-ferromagnetic type order while the DM-type couplings favor
spiral type order, the interplay between them can drive the system
into various exotic magnetically ordered states.

\begin{figure}
\includegraphics[width=3.1in]{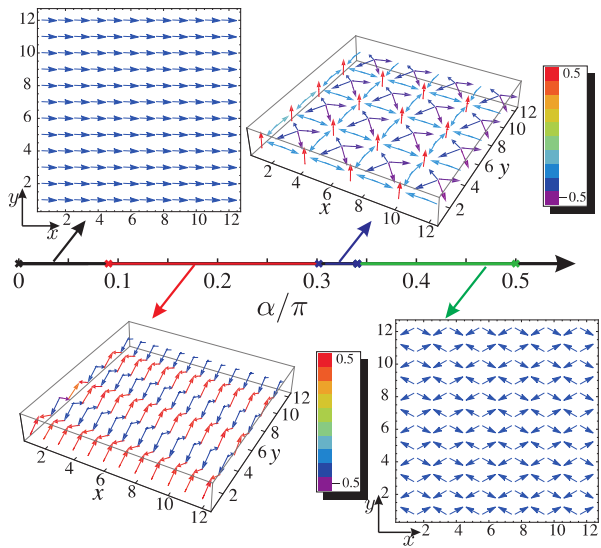} \caption{(color online). Ground state phase diagram in the deep Mott-insulator
regime ($t=0.001U$) in terms of Rashba-type SO coupling parameter
$\alpha$ at a fixed interaction ratio $\lambda=0.8$ obtained by
simulations on a $12\times12$ square lattice. Four types of phases,
i.e. $xy$-ferromagnetic (black interval), spiral (red interval),
$3\times3$ skyrmion lattice (blue interval), and $2\times2$ spin-vortex
(green interval) phases are found. The 2D vector plots of the $xy$-ferromagnetic
and the $2\times2$ spin-vortex phases correspond to the $xy$-magnetization
distribution, since these two phases are co-planar. The 3D vector
plots show the magnetization vector in 3D space, where the color of
the arrow denotes the magnetization in $z$-direction.\label{Flo:phase_digram_Mott_regime}}
\end{figure}

Instead of studying the effective exchange Hamiltonian (\ref{eq:H_eff}),
we directly simulate the original lattice boson Hamiltonian (\ref{eq:Hamiltonian})
within BDMFT. Our results in the deep MI regime are summarized in
Fig.~\ref{Flo:phase_digram_Mott_regime}, which shows a phase diagram
of the system at a fixed small hopping $t=0.001U$ and an interaction
ratio $\lambda=0.8$. When the SO coupling is weak ($|\alpha|\ll1$)
the system is dominated by the H-type exchange interactions. For $\lambda<1$
the spin-exchange in the $xy$-plane is larger than the one in the
$z$-direction and ferromagnetic, hence the system favors an $xy-$ferromagnetic
phase as expected. Upon further increasing the SO coupling, the DM-type
interaction begins to play a role and drives a transition from the
$xy-$ferromagnet to the spiral magnetic phases (red interval in the
phase diagram in Fig.~\ref{Flo:phase_digram_Mott_regime}). Due to
the limitation of finite system size in our simulations, we are not
able to resolve possible incommensurate spiral order. For simulations
on a $12\times12$ lattice presented in this paper, we are able to
resolve commensurate spiral order of spatial period $12,6$ and $4$
lattice sites, denoted as spiral-$12$, spiral-$6$ and spiral-$4$
for simplicity. To be more specific, in the simulations on a $12\times12$
lattice, we observe a sequence of transitions from spiral-$12$ to
spiral-6 to spiral-4 upon increasing SO coupling strength $\alpha$.
Further increasing $\alpha$ drives the system from the spiral phase
to a $3\times3$ skyrmion phase (blue interval in the phase diagram
in Fig.~\ref{Flo:phase_digram_Mott_regime}). 

Further increasing $\alpha$ drives another transition from the $ $$3\times3$
skyrmion phase to a $2\times2$ spin vortex whose existence can be
attributed to the spin-flip hopping process due to SO coupling. To
make this point more specific, let us focus on the limit $|\alpha-\pi/2|\ll1$.
From the explicit form of the hopping matrix element $\mathcal{R}_{ij}$,
we can see that the conventional spin-conserving hopping is strongly
suppressed in this case, i.e. particles on neighboring sites are mainly
coupled by spin-flip hopping processes arising from SO coupling. This
gives rise to an unconventional H-type interaction in the effective
spin Hamiltonian with a spatially-isotropic antiferromagnetic interaction
in $z$-direction, and exchange interactions which are anisotropic
both in the spatial directions and spin orientations in the $xy$-plane.
Since the H-type exchange couplings dominate over the DM-type ones
in this regime, they determine the magnetic order. For $\lambda<1$
the spin interaction in the $xy$-plane is larger than the one in
the $z$-direction, hence the system favors a co-planar magnetic order.
As we can see from the explicit form of $J_{\delta}^{a}$ the exchange
for the $x$ component is antiferromagnetic along the spatial $x$
direction, but ferromagnetic along the $y$ direction. This indicates
that the $x$-component of the spins favors to align parallel along
the (spatial) $y$ direction, while anti-parallel along the $x$ direction.
Similarly, one can see that the $y$-component of the spins favors
to align parallel along the $x$ direction but anti-parallel along
the $y$ direction. As a result, this unconventional H-type interaction
gives rise to a $2\times2$ spin vortex configuration in the $xy$-plane,
as shown in the right bottom part of Fig.~\ref{Flo:phase_digram_Mott_regime}.

\begin{figure}
\includegraphics[width=3.3in]{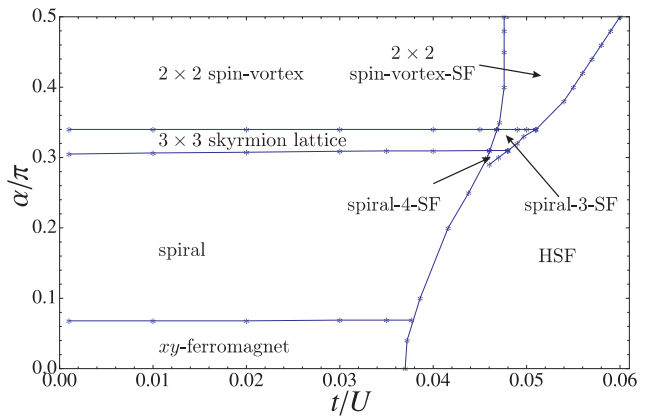} \caption{(color online). Ground state phase diagram in terms of SO coupling
strength $\alpha$ and hopping amplitude $t$ at a fixed interaction
ratio $\lambda=0.8$, obtained by simulations on a $12\times12$ square
lattice. There are four magnetic phases in the Mott-insulator regime:
$xy$-ferromagnet, spiral phase, $3\times3$ skyrmion lattice phase
and $2\times2$ spin-vortex phase. In the SF regime three phases exist,
namely the homogenous superfluid (HSF), a superfluid with $2\times2$
spin-vortex magnetic texture ($2\times2$ spin-vortex-SF), and a superfluid
with spiral magnetic texture with spatial period of 3 lattice sites
(spiral-3-SF).\label{Flo:phase_digram_alpha_t}}
\end{figure}

Although we thus confirm numerically that the exchange Hamiltonian
Eq. (\ref{eq:H_eff}) is a good description in the deep MI regime,
due to the perturbative nature of the effective spin model, it is
important to investigate whether these exotic magnetic phases are
stable against quantum fluctuations for large hopping $t/U$. For
this purpose, we calculate the full $\alpha-t$ phase diagram within
BDMFT, as shown in Fig.~\ref{Flo:phase_digram_alpha_t}. We find
that, although the validity of the effective spin model (\ref{eq:H_eff})
is restricted to the deep MI regime, the exotic spin-textures are
robust throughout the whole Mott regime up to small shifts of the
boundaries between the different phases. These results are of practical
importance in experiments aiming at observing these exotic magnetic
phases, since the effective exchange interaction strengths are proportional
to $t^{2}$, yielding higher critical temperatures ($\sim k_{B}^{-1}t^{2}/V$
with $k_{B}$ being the Boltzmann constant) of the magnetic phases
for larger values of $t$. 

\begin{figure*}
\includegraphics[width=6in]{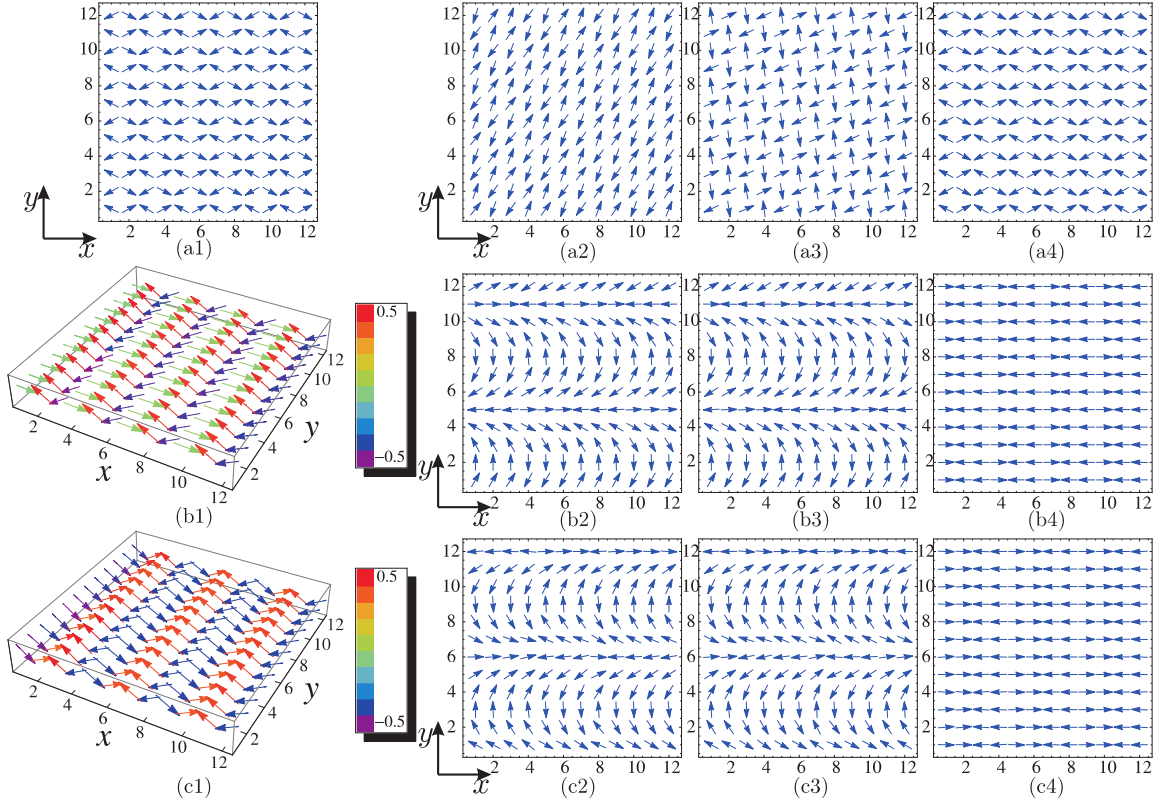} \caption{(color online). Typical superfluid phases with non-uniform magnetic
order. From up to down, different rows of plots correspond to $2\times2$
spin-vortex SF phase, spiral-3 SF phase and spiral-4 SF phase respectively.
From left to right different columns corresponds to the real-space
distribution of magnetization, $\arg(\langle a_{i\uparrow}\rangle)$,
$\arg(\langle a_{i\downarrow}\rangle)$ and $\arg\left(\langle a_{i\uparrow}\rangle^{*}\langle a_{i\downarrow}\rangle\right)$.
For the $2\times2$ spin-vortex SF phase since the magnetization is
co-planar, only a 2D vector plot is shown. The 3D vector plots correspond
to the magnetization vector in 3D space and the color of the arrow
denotes the magnetization in $z$-direction. \label{Flo:Magnetic_SF_configuration}}
\end{figure*}

\subsection{MI-SF transition and exotic superfluid phases with magnetic textures}

As it is well known, the MI-SF transition occurs when the bandwidth
($\sim8t$ for square lattice) is comparable to the interaction strength
$U$. However, in Fig.~\ref{Flo:phase_digram_alpha_t} we find an
$\alpha$-dependence of the SF-MI transition boundary. This can be
qualitatively understood within the slave-boson picture \cite{SOC_BOS_lat_1}.
The spinful lattice boson annihilation (creation) operator $\hat{a}_{i\sigma}$($\hat{a}_{i\sigma}^{\dagger}$)
in the original lattice boson Hamiltonian (\ref{eq:Hamiltonian}),
can be decomposed into the bosonic annihilation (creation) operators
of ``spinon'' $\hat{f}_{i\sigma}(\hat{f}_{i\sigma}^{\dagger})$
and ``chargon'' $\hat{c}_{i}(\hat{c}_{i}^{\dagger})$, with $\hat{a}_{i\sigma}=\frac{1}{\sqrt{\hat{n}_{i}}}\hat{c}_{i}\hat{f}_{i\sigma}$,
$\hat{n}_{i}=\hat{c}_{i}^{\dagger}\hat{c}_{i}$, and the constraint
$ $$\sum_{\sigma}\hat{f}_{i\sigma}^{\dagger}\hat{f}_{i\sigma}=\hat{c}_{i}^{\dagger}\hat{c}_{i}$
projected to the physical Hilbert space. Then, the lattice-boson Hamiltonian
(\ref{eq:Hamiltonian}) assumes another representation, denoted as
$H_{\mathrm{SC}}$, in terms of annihilation (creation) operators
of spinon and chargon, 
\begin{eqnarray}
H_{\mathrm{SC}} & = & -t(\frac{1}{\sqrt{\hat{n}_{i}\hat{n}_{j}}}\hat{c}_{i}^{\dagger}\hat{c}_{j}\hat{\Phi}_{i}^{\dagger}\mathcal{R}_{ij}\hat{\Phi}_{j}+h.c.)\label{eq:Hamiltonian_in_Spinon-Chargon-Basis}\\
 &  & -\mu\sum_{i}\hat{\Phi}_{i}^{\dagger}\hat{\Phi}_{i}-\sum_{i}\mu_{i}^{c}(\hat{c}_{i}^{\dagger}\hat{c}_{i}-\hat{\Phi}_{i}^{\dagger}\hat{\Phi}_{i})\nonumber \\
 &  & +\frac{U}{2}\sum_{i}(\hat{n}_{f,i,\uparrow}^{2}+\hat{n}_{f,i,\downarrow}^{2}+2\lambda\hat{n}_{f,i,\uparrow}\hat{n}_{f,i,\downarrow}),\nonumber 
\end{eqnarray}
where $\hat{\Phi}_{i}^{\dagger}\equiv(\hat{f}_{i\uparrow}^{\dagger},\hat{f}_{i\downarrow}^{\dagger})$
and $\mu_{i}^{c}$ is introduced to keep the local constraint satisfied
on average $\langle\sum_{\sigma}\hat{f}_{i\sigma}^{\dagger}\hat{f}_{i\sigma}\rangle=\langle\hat{c}_{i}^{\dagger}\hat{c}_{i}\rangle$.
By making a mean-field approximation to the spinon, i.e. $\hat{\Phi}_{i}\rightarrow\Phi_{i}$
with $\Phi_{i}$'s being $c$-number valued vectors, and keeping only
the leading order contribution of the ``spinon'' to the physics
of the chargon, the effective Hamiltonian of the chargon degree of
freedom, denoted as $H_{\mathrm{C}}^{\mathrm{eff}}$, assumes the
form of the conventional Bose-Hubbard model 
\begin{equation}
H_{\mathrm{C}}^{\mathrm{eff}}=-\sum_{\langle i,j\rangle}(\tilde{t}_{ij}\hat{c}_{i}^{\dagger}\hat{c}_{j}+h.c.)+\frac{U}{2}\sum_{i}\hat{c}_{i}^{\dagger}\hat{c}_{i}^{\dagger}\hat{c}_{i}\hat{c}_{i}.\label{eq:Effective_Hamiltonian_Chargons}
\end{equation}
where $\tilde{t}_{ij}=t\Phi_{i}^{\dagger}\mathcal{R}_{ij}\Phi_{j}$
can be regarded as the ``spinon-renormalized'' hopping amplitude
of the chargon. Since the MI-SF transition is driven by the chargons,
from Eq. (\ref{eq:Effective_Hamiltonian_Chargons}) we directly see
that different mean-field configurations of the spinon, which correspond
to different magnetic configurations, give rise to different renormalized
chargon hopping amplitudes $\tilde{t}_{ij}$, and hence make the MI-SF
transition boundary $\alpha$-dependent.

From the phase diagram in Fig.~\ref{Flo:phase_digram_alpha_t}, we
see that SF phases with magnetic textures emerge near the MI-SF transition.
In particular, we observe a new spin-spiral magnetic texture with
spatial period of $3$ (spiral-3) in the SF phase near the MI-SF transition
{[}see Fig.~\ref{Flo:Magnetic_SF_configuration}(b1){]}. A qualitative
picture is that the chargon degree of freedom undergoes drastic changes
across the MI-SF transition, and the feedback of the chargon to the
spinon degree of freedom modifies the magnetic textures in the SF
phase pronouncedly. From our simulation results, we observe that the
phase angle distribution of the superfluid order parameters has complex
structures {[}see Fig.~\ref{Flo:Magnetic_SF_configuration}(a2),
(b2), (c2), (a3), (b3), and (c3){]}. Interestingly, for the superfluid
with $2\times2$ spin-vortex magnetic structure, superfluid vortices
are observed as well {[}see Fig.~\ref{Flo:Magnetic_SF_configuration}(a3){]}.
Moreover, it is interesting to see that the relative phase $\Delta\theta_{i}\equiv\theta_{i\downarrow}-\theta_{i\uparrow}$
between the two superfluid components has a simple texture which coincides
with the corresponding magnetic texture on the $x-y$ plane {[}see
Fig.~\ref{Flo:Magnetic_SF_configuration}(a4), (b4), and (c4){]}.
Qualitatively, this can be understood within the slave-boson picture.
The condensate order parameter in the slave-boson representation reads
$\phi_{\sigma}\equiv\langle a_{i\sigma}\rangle=\langle c_{i}f_{i\sigma}\rangle$.
Assuming that the correlations between chargon and spinon are small,
we approximate $\phi_{i\sigma}\approx\langle c_{i}\rangle\langle f_{i\sigma}\rangle$.
Then the relative phase between the two superfluid components is given
by $\arg(\phi_{i\uparrow}^{*}\phi_{i\downarrow})=\arg(\langle f_{i\uparrow}\rangle^{*}\langle f_{i\downarrow}\rangle)$,
together with the fact that the magnetization on the $x-y$ plane
is given by $\langle S_{i}^{x}\rangle=\Phi_{i}^{\dagger}\sigma_{x}\Phi_{i}/2=\left|\langle f_{i\uparrow}\rangle\langle f_{i\downarrow}\rangle\right|\cos\left(\arg(\langle f_{i\uparrow}\rangle^{*}\langle f_{i\downarrow}\rangle)\right)$
and $\langle S_{i}^{y}\rangle=\Phi_{i}^{\dagger}\sigma_{y}\Phi_{i}/2=\left|\langle f_{i\uparrow}\rangle\langle f_{i\downarrow}\rangle\right|\sin\left(\arg(\langle f_{i\uparrow}\rangle^{*}\langle f_{i\downarrow}\rangle)\right)$,
we see that the texture of the relative phase between the two superfluid
components is indeed the same as the magnetic texture in the $x-y$
plane.

\section{Conclusion}

We have investigated ground state properties of Bose-Bose mixtures
with Rashba-type spin-orbit (SO) coupling loaded in a square optical
lattice, within the non-perturbative theoretical framework of BDMFT.
The system shows rich spin physics in both the Mott-insulator and
the superfluid regime. We found that the exotic magnetic phases are
robust in the whole MI regime. In the proximity of the MI-SF transition,
exotic superfluid phases with magnetic textures arise as a result
of the interplay between onsite interaction and SO coupling. In particular,
we identify a new spin-spiral magnetic texture with spatial period
$3$ on the SF side near the MI-SF transition. These results are of
practical importance for experimental observation of these exotic
spin texture and novel SF phases.
\begin{acknowledgments}
L. H. thanks X. Zhang, D. Cocks, M. Buchhold, S. Diehl, R. Li, and
Q. Sun for useful discussions. This work was supported by the Deutsche
Forschungsgemeinschaft DFG via Sonderforschungsbereich SFB/TR 49 and
Forschergruppe FOR 801. A. C. Ji was supported by NCET and NSFC under
grants No. 11474205.\end{acknowledgments}

\end{document}